\documentclass[preprint]{revtex4}

\usepackage{graphicx}

\begin{document}

\title{Force-Free Electrodynamics of Pulsars}

\author{Andrei Gruzinov}
 
\affiliation{Center for Cosmology and Particle Physics, Department of Physics, New York University, NY 10003}

\date{April 17, 2006}

\begin{abstract}

We show that FFE (Force-Free Electrodynamics) describes pulsar magnetospheres -- even when radiation and particle wind power are comparable to the spin-down power. At the same time, we show that FFE is insufficient for calculating pulsar magnetospheres. This is because FFE admits a large family of stationary solutions with different currents flowing in the closed line region. To choose the actual solution one needs a model of current generation which goes beyond pure FFE. 

We calculate several FFE magnetospheres for the aligned rotator. The Poynting power of these solutions fills the range $c^{-3}\mu ^2\Omega^4~< ~L~ <~ 0.67 c^{-3}\mu ^2\Omega^4(c/\Omega R_s)^2$, for angular velocity $\Omega$ and magnetic field which is a pure dipole $\mu$ on the surface of the star of radius $R_s$.

Anomalous braking indices of young pulsars might be explained by currents flowing in the closed-line region.

\end{abstract}

\maketitle

\section{Introduction}

It might appear that pulsar magnetosphere and spin-down power have finally been calculated. Both aligned rotator (pulsar with small spin-dipole inclination) \cite{goldreich, scharlemant, michel, contopoulos, gruzinov1, komissarov, mckinney, spitkovsky, timokhin, bucciantini} and inclined rotator (arbitrary spin-dipole angle) \cite{gruzinov2, spitkovsky} have been solved. Pulsar magnetospheres were calculated in the framework of force-free electrodynamics (FFE, \S2). Magnetohydrodynamics (MHD) codes were also used, but we will argue that MHD applies to pulsars only inasmuch as it reduces to FFE. 

While we do think that FFE is applicable to pulsars (\S2), we also show that FFE description of pulsars is ambiguous. In particular, we show (\S3) that FFE can only give lower and upper bounds for the pulsar Poynting power:
\begin{equation}\label{bound}
{\mu ^2\Omega^4\over c^3}~\lesssim ~L~\lesssim ~{\mu ^2\Omega^4\over c^3}\left( {c\over \Omega R_s}\right)^2,
\end{equation}
where $R_s$ is the radius of the star.

Observations might actually favor replacing the standard power formula, $L\sim c^{-3}\mu ^2\Omega ^4$, with the bounds (\ref{bound}). Anomalous braking indices of young pulsars are best explained by the spin-down power $L \propto \Omega^{4-\alpha}$ \cite{gruzinov3}. This spin-down power obtains if we assume that Poynting and spin-down power are comparable, and also assume that $L$ is proportional to some power of $\Omega$ which falls into the range allowed by FFE bounds. Then we get the spin-down power $L \sim c^{-3}\mu ^2\Omega^4(c/\Omega R_s)^{\alpha }$, with $0<\alpha <2$, in agreement with observations.

\section{Force-Free Electrodynamics of Pulsars}

Here we show that FFE describes large-scale electromagnetic field of pulsars, even when particle or radiation power are comparable to the large-scale Poynting power.

FFE postulates that electromagnetic field satisfies Maxwell equations,
\begin{equation}\label{ffee}
\partial _t{\bf B}=-\nabla \times {\bf E},~~~\partial _t{\bf E}=\nabla \times {\bf B}-{\bf j},
\end{equation}
with the current 
\begin{equation}\label{ohm}
{\bf j}={({\bf B}\cdot \nabla \times {\bf B}-{\bf E}\cdot \nabla \times {\bf E}){\bf B}+(\nabla \cdot {\bf E}){\bf E}\times {\bf B} \over B^2}.
\end{equation}
Initial condition ${\bf E}\cdot {\bf B}=0$ is assumed, and then this condition is sustained by FFE. 

The strange-looking Ohm's law (\ref{ohm}) is designed \cite{gruzinov4} in such a way that force-free condition, 
\begin{equation}\label{ffree}
\rho {\bf E}+{\bf j}\times {\bf B}=0
\end{equation}
(with $\rho \equiv \nabla \cdot {\bf E}$) is always satisfied. We will explain physical meaning of the FFE Ohm's law below.

Standard derivation of FFE postulates that particle inertia is negligible compared to characteristic values of electromagnetic forces. Then the electromagnetic force density should vanish, as given by the force-free condition (\ref{ffree}). One then finds an expression for the current ${\bf j}$ which sustains this condition as the fields evolve according to Maxwell equations (\ref{ffee}).

The force-free condition is Lorentz-invariant: $F^{\mu \nu}j_\nu=0$, where $F^{\mu \nu}$ is the electromagnetic field tensor and $j^\mu$ is the 4-current. Since divergence of the electromagnetic energy-momentum tensor is $\partial _\nu T^{\mu \nu}=-F^{\mu \nu}j_\nu$, FFE should apply if energy-momentum tensor is dominated by the electromagnetic part. 

But this condition -- negligible energy-momentum tensor of particles -- might be too restrictive. For example, X-ray luminosity of Crab-like pulsars can be up to 10\% of spin-down power \cite{possenti}. Since X-ray energy is coming from particles, the particle power might be actually comparable or even dominate the spin-down power. If so, the {\it derivation} of FFE presented above fails. However, we think that even in such cases FFE actually remains valid, because FFE equations can be derived without postulating small particle inertia.

We start with a definition. Consider arbitrary electromagnetic field. At each event (space-time event), one can find a {\it good } reference frame, where the electric field ${\bf E}$ is parallel to the magnetic field ${\bf B}$. In fact, there is a whole one-parameter family of good frames, obtained by boosting a good frame along common direction of ${\bf E}$ and ${\bf B}$. Which good frame is chosen is irrelevant for our purposes. 

FFE follows from two assumption. In a good frame: (i) electric field vanishes, (ii) current flows along ${\bf B}$. These conditions obviously give force freedom (\ref{ffree}) in a good frame, and therefore in any frame, and thus these conditions are equivalent to FFE. The physical meaning of FFE Ohm's law is most clear in a good frame: the current flows along ${\bf B}$, and the magnitude of the current is chosen in such way as to sustain zero electric field. We will now argue that these conditions should indeed be satisfied, at least in Crab-like and Vela-like pulsars.

 The first condition means that electric field component along magnetic field at radius $r$ should be much smaller than the characteristic value $\sim (\Omega r/c)B$. Consider young pulsar with nominal surface field $B_s\sim 3\times 10^{12}$ G and angular velocity $\Omega \sim 100$ s$^{-1}$. Then electron-positron pair creation avalanche (\cite{ruderman}: electrons and positrons produce gamma-rays by curvature radiation, gamma rays produce pairs on magnetic field) discharge the parallel component of electric field at distances $\lesssim 600$km. At larger distances magnetic field becomes too weak for pair production. The light cylinder radius is larger, $c/\Omega =3000$km, and one might worry about shortening the field near the light cylinder. It is thought, however, that avalanches are producing dense outflows of electron-positron plasma \cite{ruderman}. With the electron-positron plasma density well above Goldreich-Julian density, the parallel component of the electric field should shorten near the light cylinder too.

The second condition -- in a good frame current flows along magnetic field -- should be easily satisfied. Fast synchrotron/cyclotron cooling puts charged particles into the lowest Landau level (cooling is fast not only for electrons and positrons, but also for ions, if there are any ions in the magnetosphere).  The magnetic dipole moment corresponding to the lowest Landau level is small and the resulting large scale perpendicular current is negligible. 

Thus: negligible particle inertia is not a must for FFE applicability. It is sufficient to have (i) easily available charges (to kill ${\bf E}$ along ${\bf B}$), (ii) strong enough magnetic field (to ensure fast perpendicular cooling). It appears that these conditions can be satisfied, at least in Vela-like and Crab-like pulsars.

\section{Stationary FFE Magnetosphere and Poynting Power}

Stationary FFE magnetosphere (electromagnetic field corotates with the star) satisfies the following equation \cite{gruzinov2} :
\begin{equation}\label{basic}
{\bf B}\times \nabla \times \left( {\bf B} + {\bf V}\times ({\bf V}\times{\bf B})\right) =0,
\end{equation}
with ${\bf V}\equiv {\bf \Omega}\times {\bf r}$ and $\nabla \cdot {\bf B}=0$. The electric field is ${\bf E}=-{\bf V}\times {\bf B}$.

\begin{figure}[b]
  \begin{center}
    \includegraphics[angle=0, width=.4\textwidth]{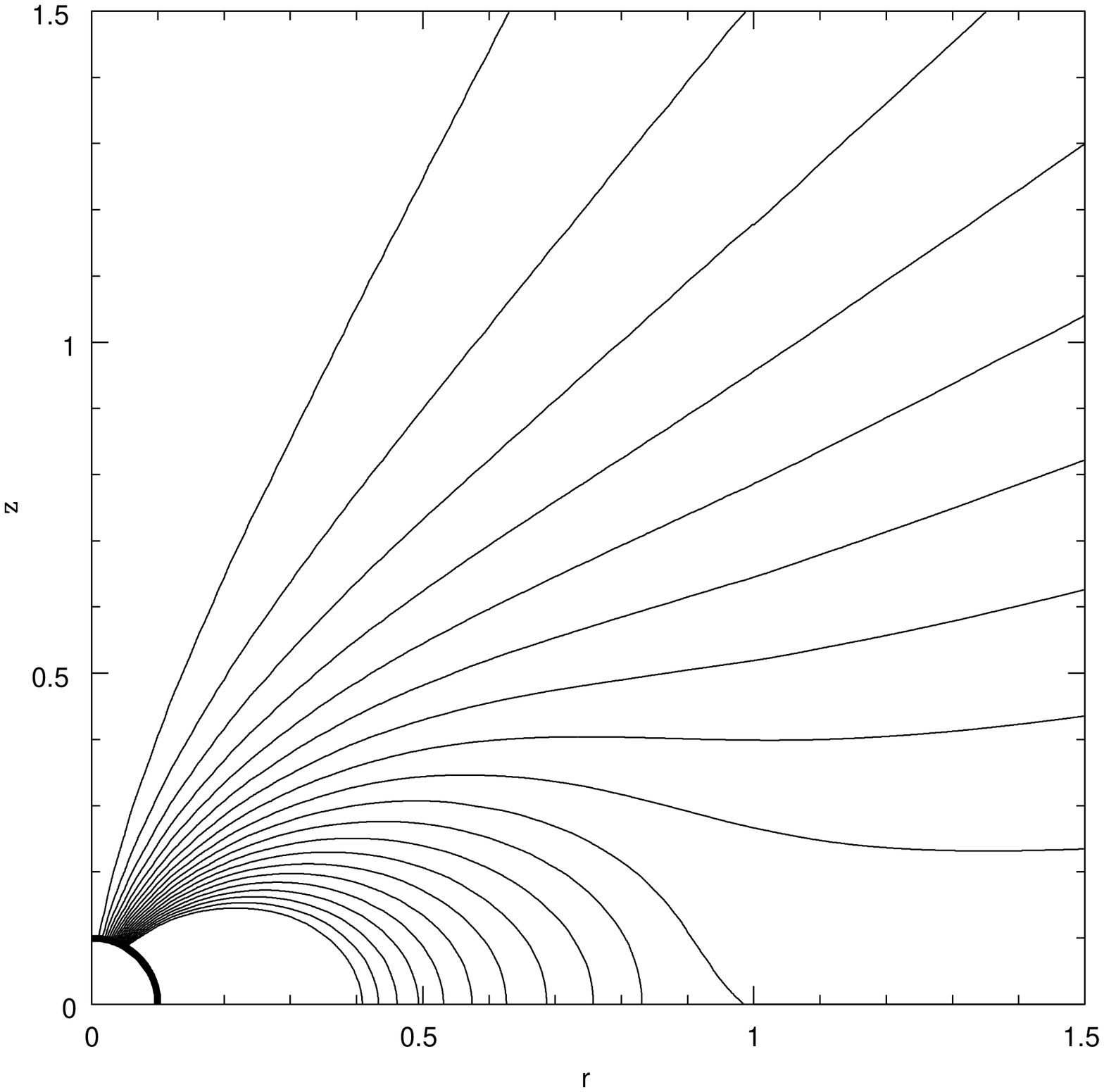}
    \includegraphics[angle=0, width=.4\textwidth]{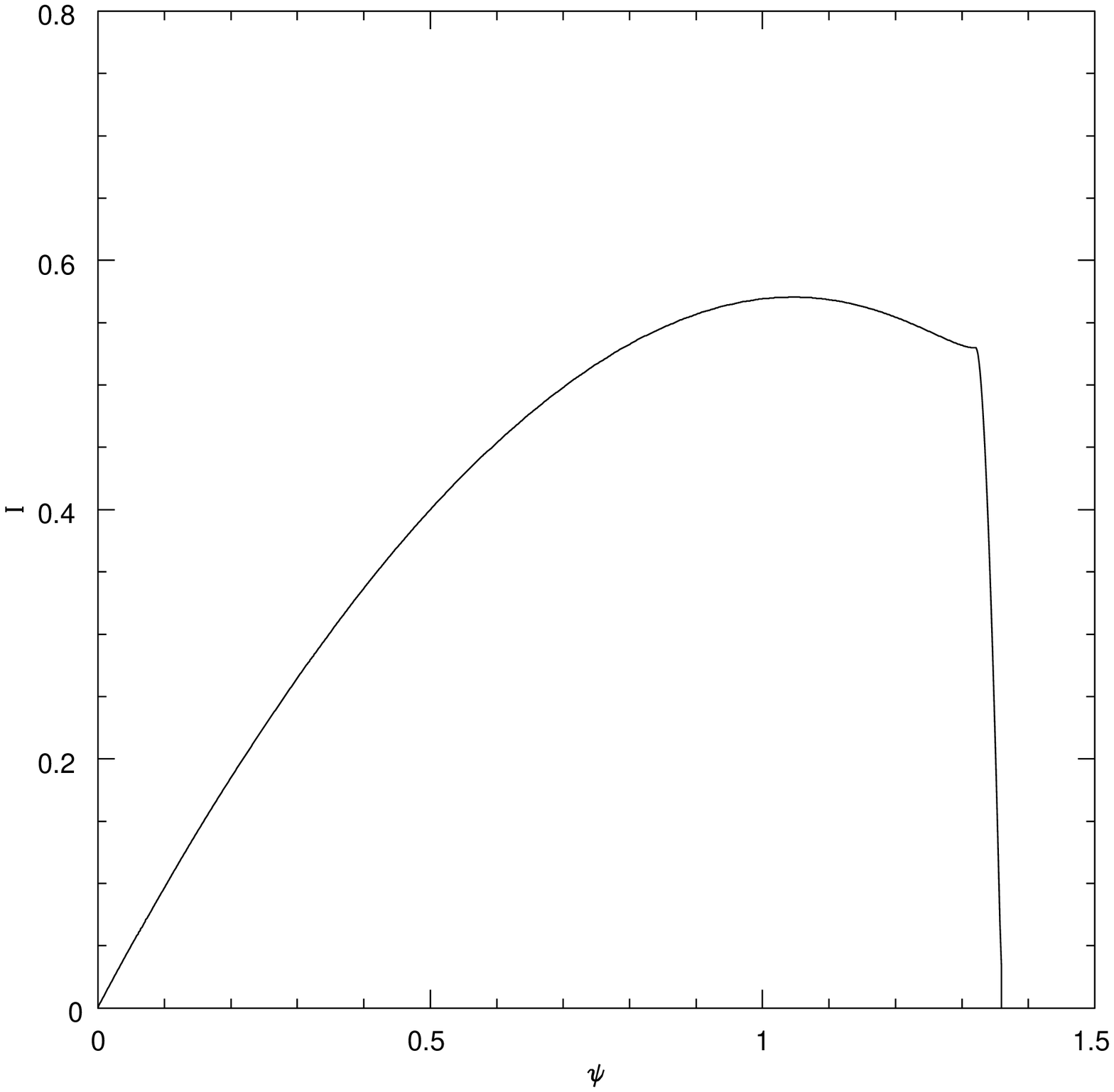}
    \caption{Standard axisymmetric pulsar. No current in the closed region. Separatrix at $\psi _0=1.32$. Power $L=1.08$. (1) $\psi$-isolines, with $\psi$-intervals of $0.1 \psi _0$. (2) Poloidal current $I(\psi )$. }
  \end{center}
\end{figure}

Solutions of the pulsar magnetosphere equation (\ref{basic}) follow from the isotopological variation variation principle: $\delta S=0$, where $S$ is proportional to electromagnetic action
\begin{equation}\label{vary}
S=\int d^3r (B^2-({\bf V}\times{\bf B})^2),
\end{equation}
and the variation of action should be calculated for isotopological perturbation of magnetic field 
\begin{equation}\label{isotop}
\delta {\bf B}=\nabla \times (\delta \vec{\xi }\times{\bf B}).
\end{equation}
In other words, action should be extremized by continuously displacing magnetic field lines. The footpoints of the lines remain fixed. The number density of footpoints is given by the normal component of magnetic field on the surface of the star $B_n(\theta , \phi)$, where  $\theta$ and $\phi$ are spherical angles.

There are magnetic field lines of two types in the stationary magnetosphere: star-infinity and star-star. The star-star lines define a mapping of the closed-line part of the stellar surface onto itself.

\begin{figure}[b]
  \begin{center}
    \includegraphics[angle=0, width=.4\textwidth]{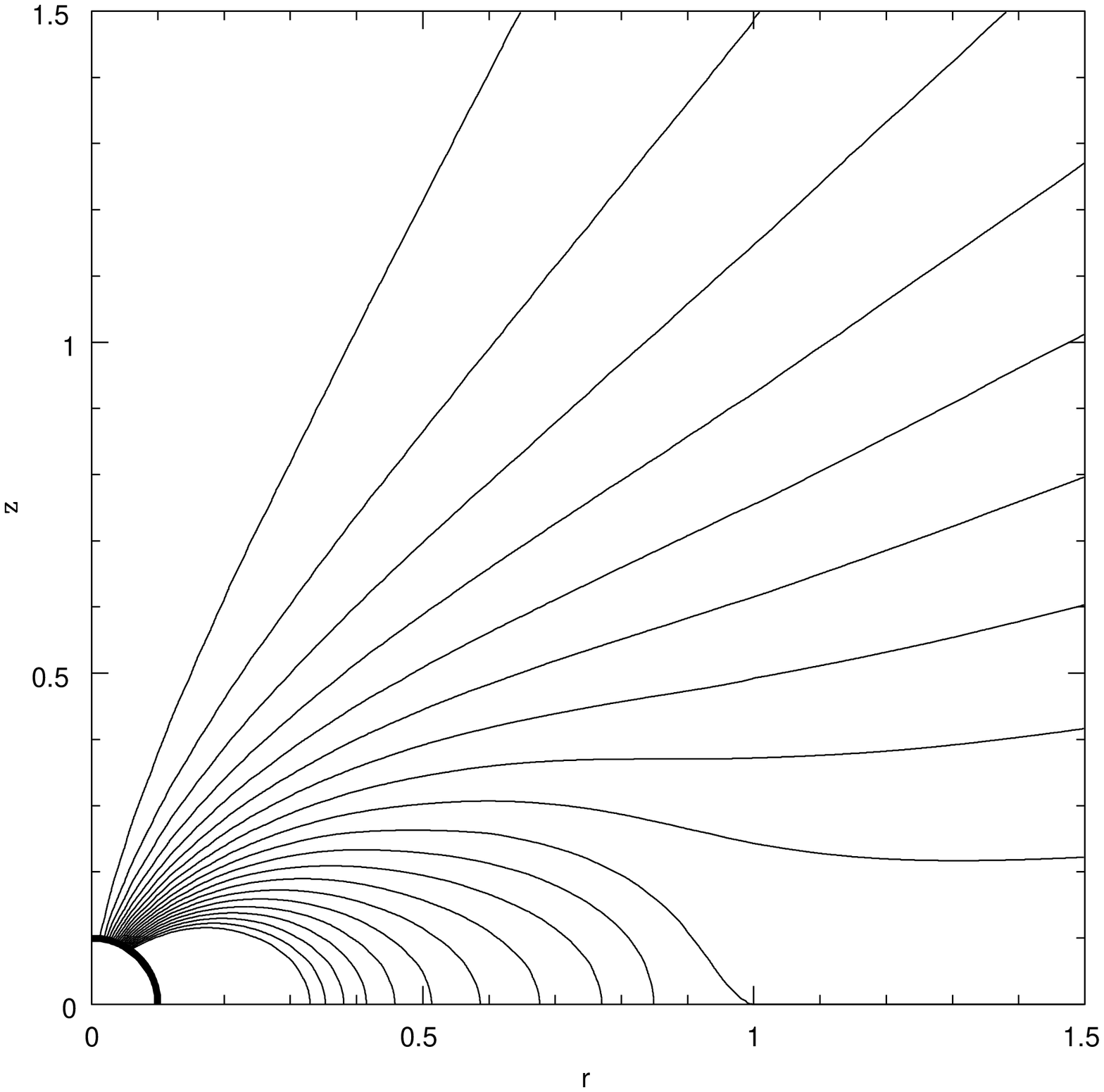}
    \includegraphics[angle=0, width=.4\textwidth]{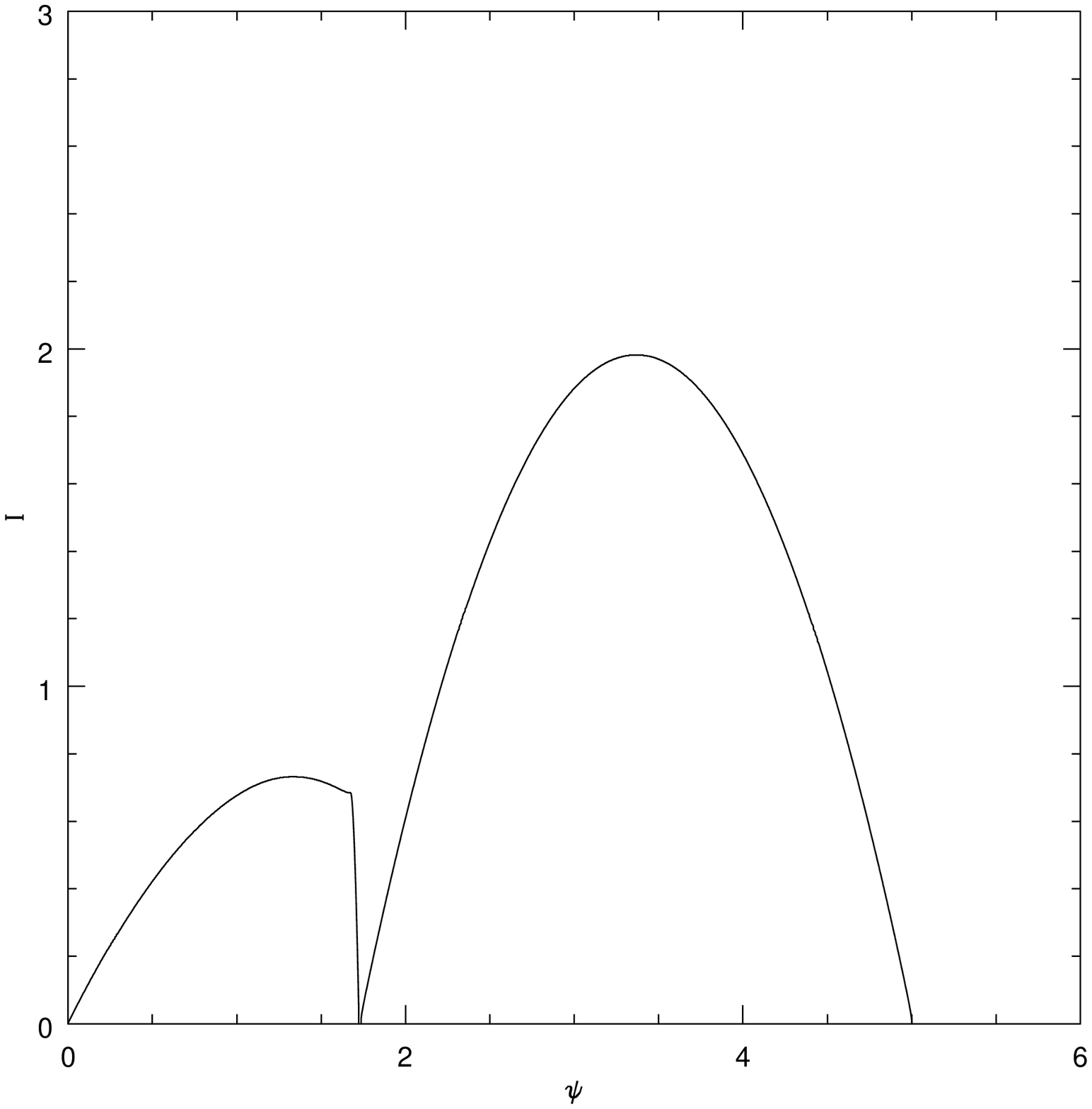}
    \caption{With current in the closed region. Separatrix at $\psi _0=1.67$. Power $L=1.75$.}
  \end{center}
\end{figure}

\begin{figure}
  \begin{center}
    \includegraphics[angle=0, width=.4\textwidth]{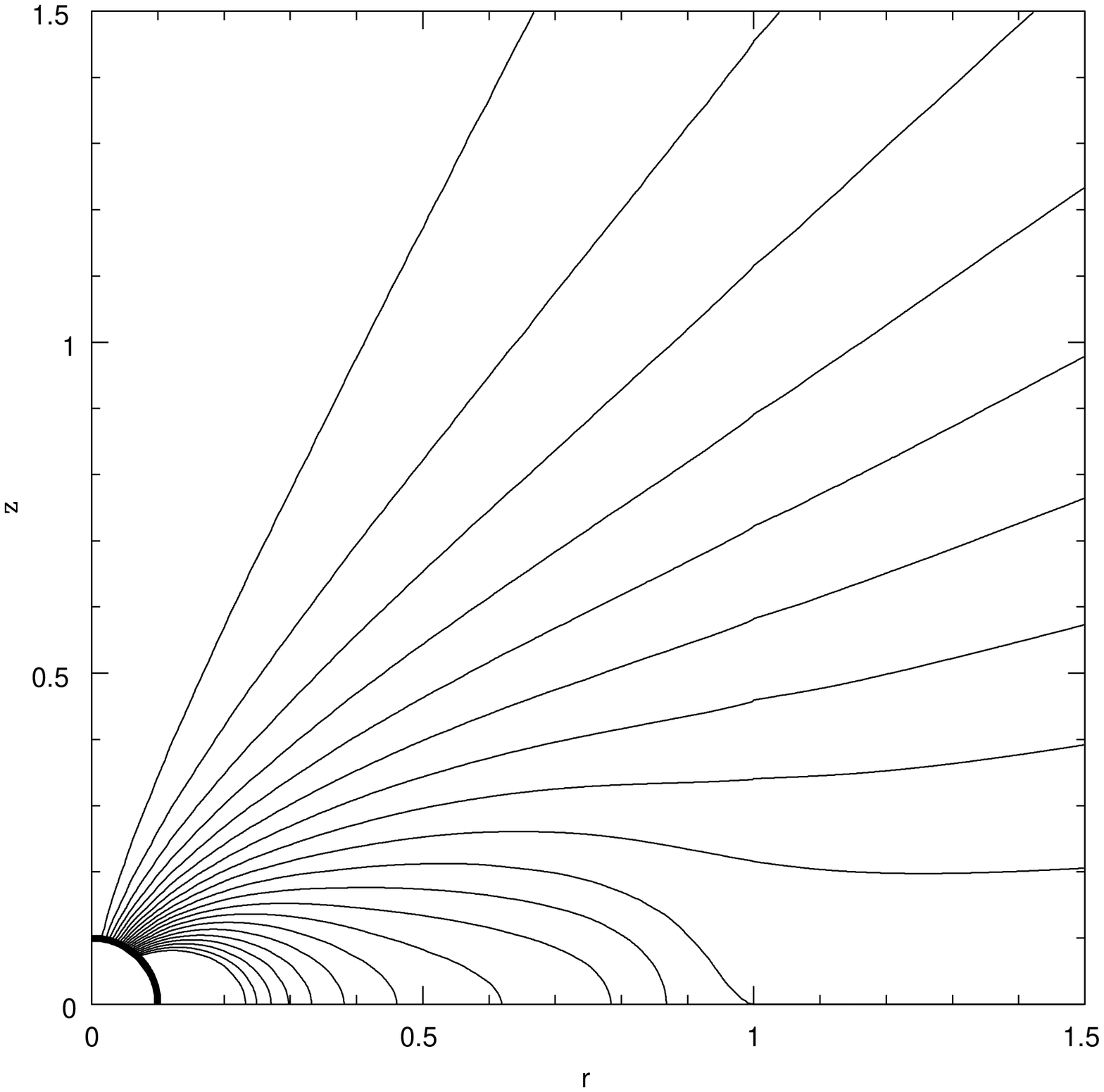}
    \includegraphics[angle=0, width=.4\textwidth]{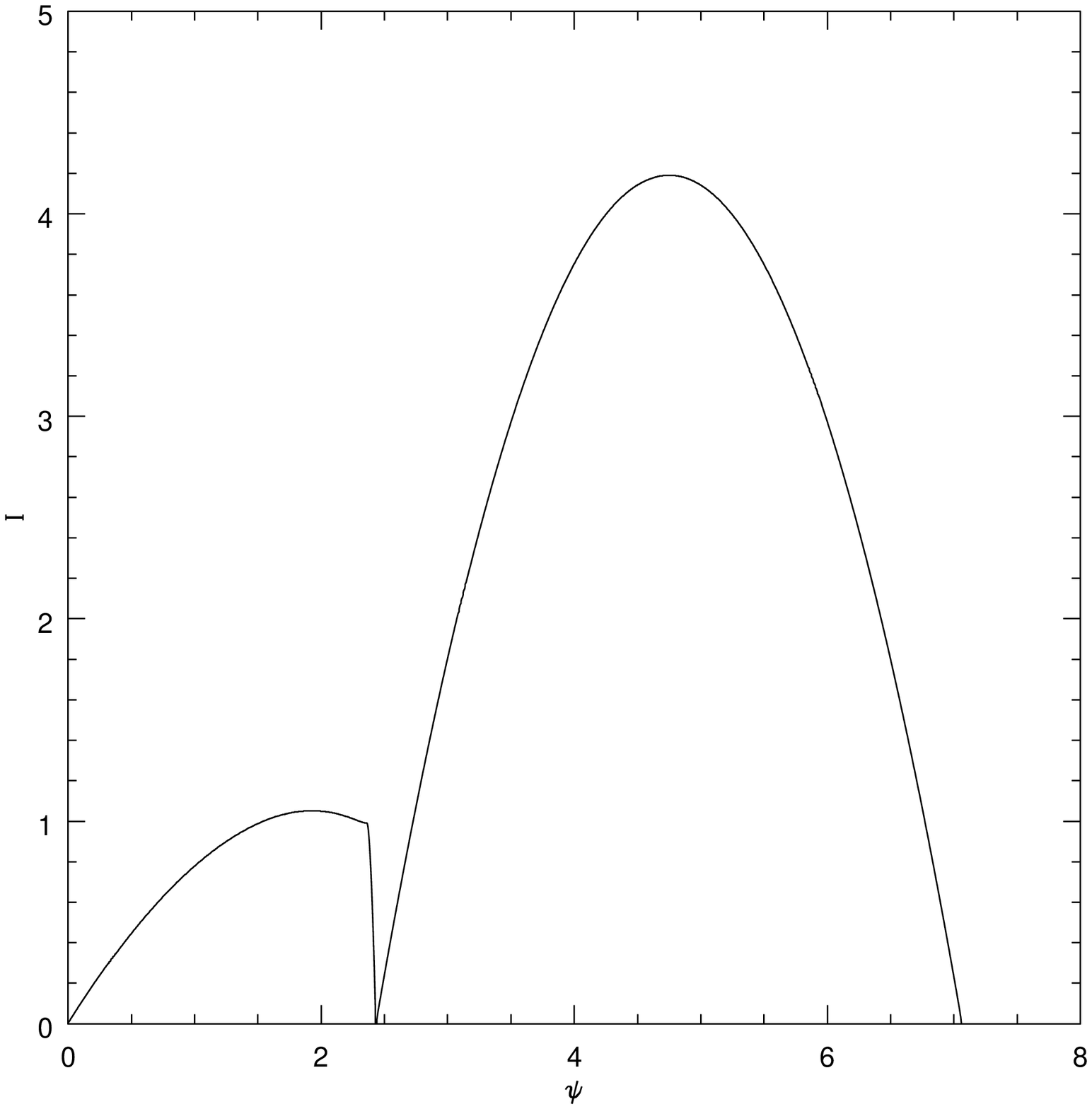}
    \caption{With current in the closed region. Separatrix at $\psi _0=2.36$. Power $L=3.65$.}
  \end{center}
\end{figure}

Stationary FFE magnetosphere can be calculated for a given (i) ${\bf \Omega}$, (ii) $B_n(\theta ,\phi )$, (iii) mapping of the closed-line part of the surface. The surface mapping should be consistent with $B_n(\theta ,\phi )$. But for a given $B_n(\theta ,\phi )$, there are infinitely many consistent mappings. Therefore there are infinitely many FFE magnetospheres with given ${\bf \Omega}$ and $B_n(\theta ,\phi )$. FFE has nothing to say about the actual mapping of the closed-line part of the stellar surface. 

Physically, different mappings correspond to different currents flowing in the closed line region. For axisymmetric pulsar, one usually postulates zero poloidal current in the closed-line region, but we see no reason why the microscopic mechanism that realizes FFE should obey this condition for a generic asymmetrical $B_n(\theta )$. The small residual parallel electric field is adjusted by the requirement that charge density be equal to Goldreich-Julian. There is no reason to expect that this adjustment process terminates at zero current. Note also that FFE generically predicts singular current layers along magnetic separatrix (surface between open and close field lines). Again there is no reason to expect that microscopic physics producing very high current densities near magnetic separatrix will not result in some finite current densities within the separatrix.

We studied the effect of the inner currents numerically. We assumed axisymmetry. For stationary axisymmetric force-free magnetic field, the magnetosphere equation (\ref{basic}) reduces to the basic equation of \cite{scharlemant} and gives magnetic field of the form 
\begin{equation}
{\bf B}={1\over r}\left( -\partial _z\psi, 2I(\psi), \partial _r\psi \right).
\end{equation}
Here $r$, $z$ are cylindrical coordinates, $\psi(r,z)/r$ is the toroidal component of the vector potential, $I(\psi )$ is the poloidal current, and $\psi(r,z)$ satisfies the following equation
\begin{equation}\label{axibasic}
(1-r^2)(\psi _{rr}+{1\over r}\psi _r+\psi _{zz})-{2\over r}\psi _r+F(\psi )=0,
\end{equation}
$F(\psi )\equiv 4I{dI\over d\psi }$. The units are $\mu =c=\Omega =1$. 

To simplify the calculation, we assumed that negligibly small north-south asymmetries launch large closed currents. The numerical procedure is just the one discovered by \cite{contopoulos}. The only difference, is that now a non-zero toroidal field (corresponding to poloidal current) exists in the closed line region. Poloidal current in the open-line region is numerically adjusted to allow smooth crossing of the light cylinder. Poloidal current in the closed-line region is chosen in an arbitrary manner. 

The results are shown in figures (1)-(4). The radius of the star is $R_s=0.1$. We have checked that we obtain legitimate FFE solutions with $B>E$. Figure (1) just repeats the standard solution. Figures (2) and (3) show solutions with increasing toroidal field in the closed region. As toroidal field increases, more lines open up and the pulsar power increases accordingly. With very large toroidal field in the closed region, one should get the split-monopole configuration shown in figure (4).

We must acknowledge that our numerical procedure for calculating stationary magnetospheres breaks down for large currents in the closed line region, and we cannot demonstrate the field configurations with $\psi _0\gtrsim 3$. However we think that such solutions do exist. The action (\ref{vary}) is positive definite inside the light cylinder and therefore it must possess a minimum for arbitrarily large twist of the initial mapping. As the lines open up, the functional (\ref{vary}) gets decreased.

A better way to argue that arbitrarily large opening of field lines can be achieved for large twists, is to consider a non-rotating FFE magnetosphere. Then equation (\ref{axibasic}) simplifies. In spherical coordinates $r=R\sin \theta$, $z=R\cos \theta$, 
\begin{equation}
\psi _{RR}+\sin \theta\left( {1\over \sin \theta }\psi _{\theta }\right) _\theta+F(\psi )=0,
\end{equation}
with selfsimilar solutions of the form
\begin{equation}
\psi \propto R^{-\alpha }, ~~~I\propto \psi ^{1+{1\over \alpha} }.
\end{equation}
We show two such solutions in Fig. (5). Note, that the twist, defined as longitude difference between footpoints of the same field line, is not too big. Assumingly, if such a star is spun up, it will generate Poynting flux proportional to magnetic energy density near the light cylinder, $L\propto R_s^{-2(1-\alpha ) }$.

We must conclude that FFE-allowed magnetospheres span a whole range between the standard magnetosphere with Poynting luminosity  $L\sim c^{-3}\mu ^2\Omega ^4$, and split-dipole magnetosphere with luminosity $L\sim c^{-3}\mu ^2\Omega ^4(c/\Omega R_s)^2$. Pure FFE cannot say which magnetosphere is actually realized. Further progress requires a model of current generation. Poloidal currents in the closed region might explain anomalous braking indices.

\begin{acknowledgments}
This work was supported by the David and Lucile Packard Foundation.
\end{acknowledgments}

\begin{figure}
  \begin{center}
    \includegraphics[angle=0, width=.4\textwidth]{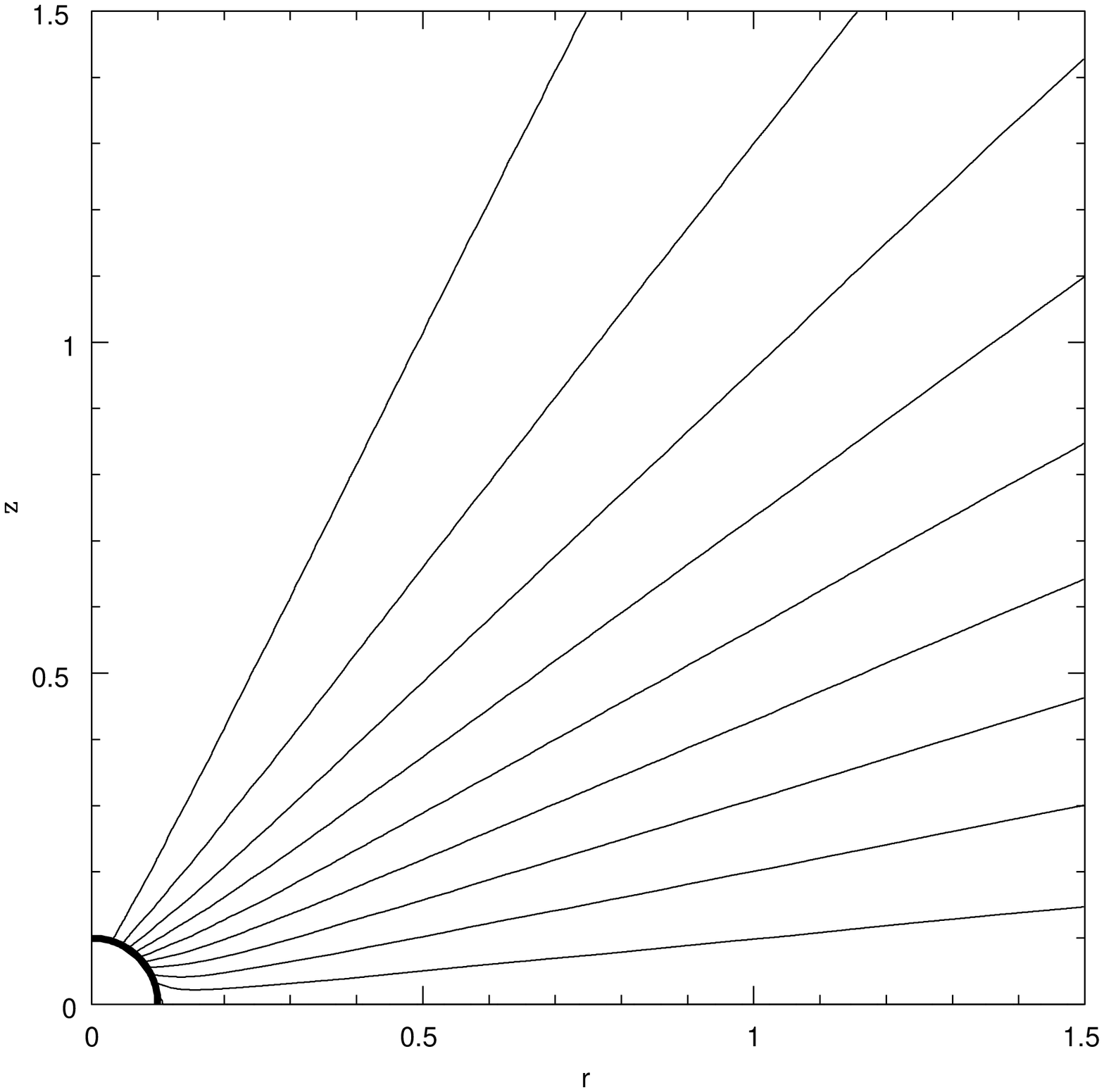}
    \includegraphics[angle=0, width=.4\textwidth]{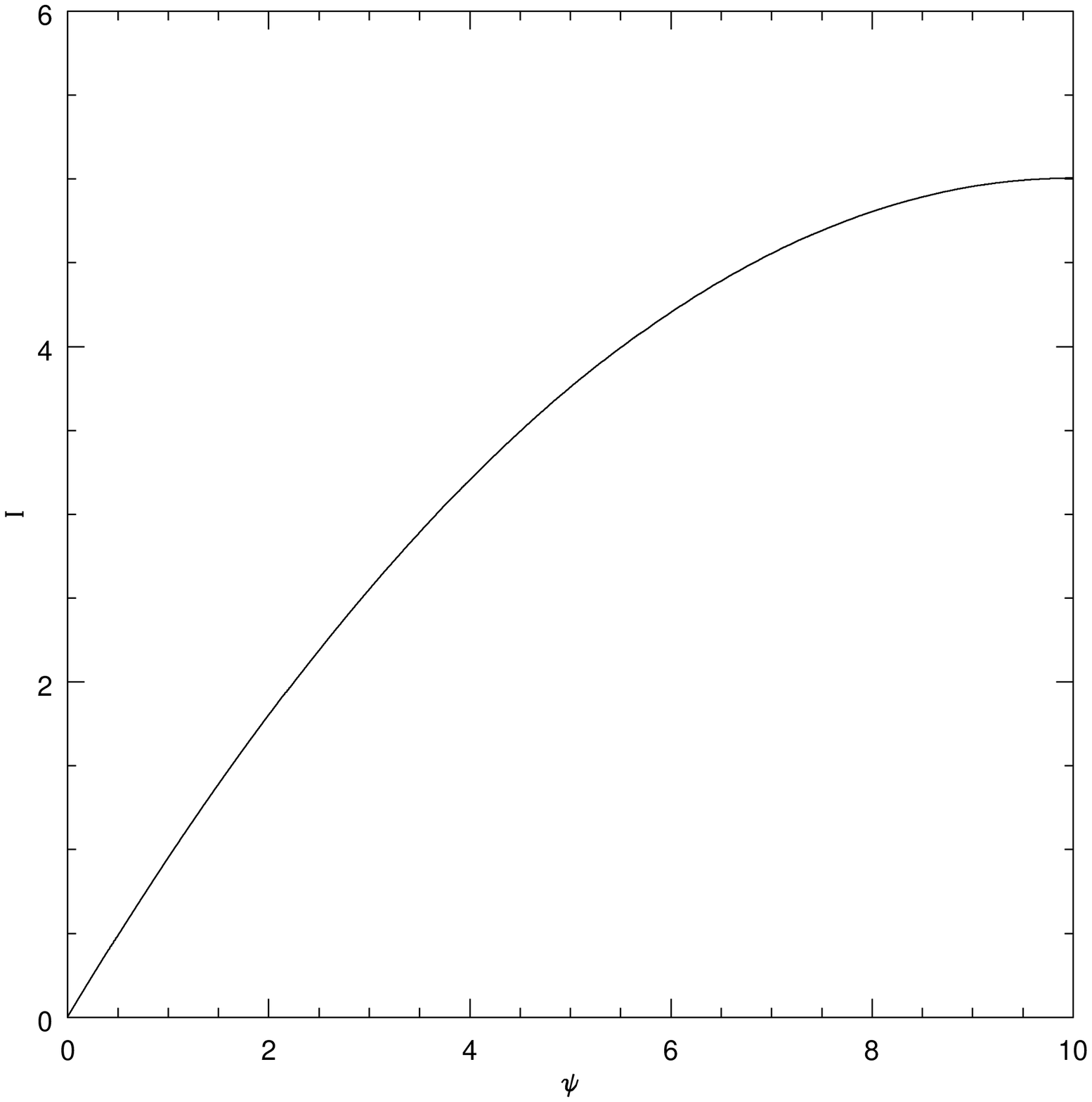}
    \caption{All lines open. Separatrix at $\psi _0=10$. Power $L=67.1$}
  \end{center}
\end{figure}

\begin{figure}[b]
  \begin{center}
    \includegraphics[angle=0, width=.4\textwidth]{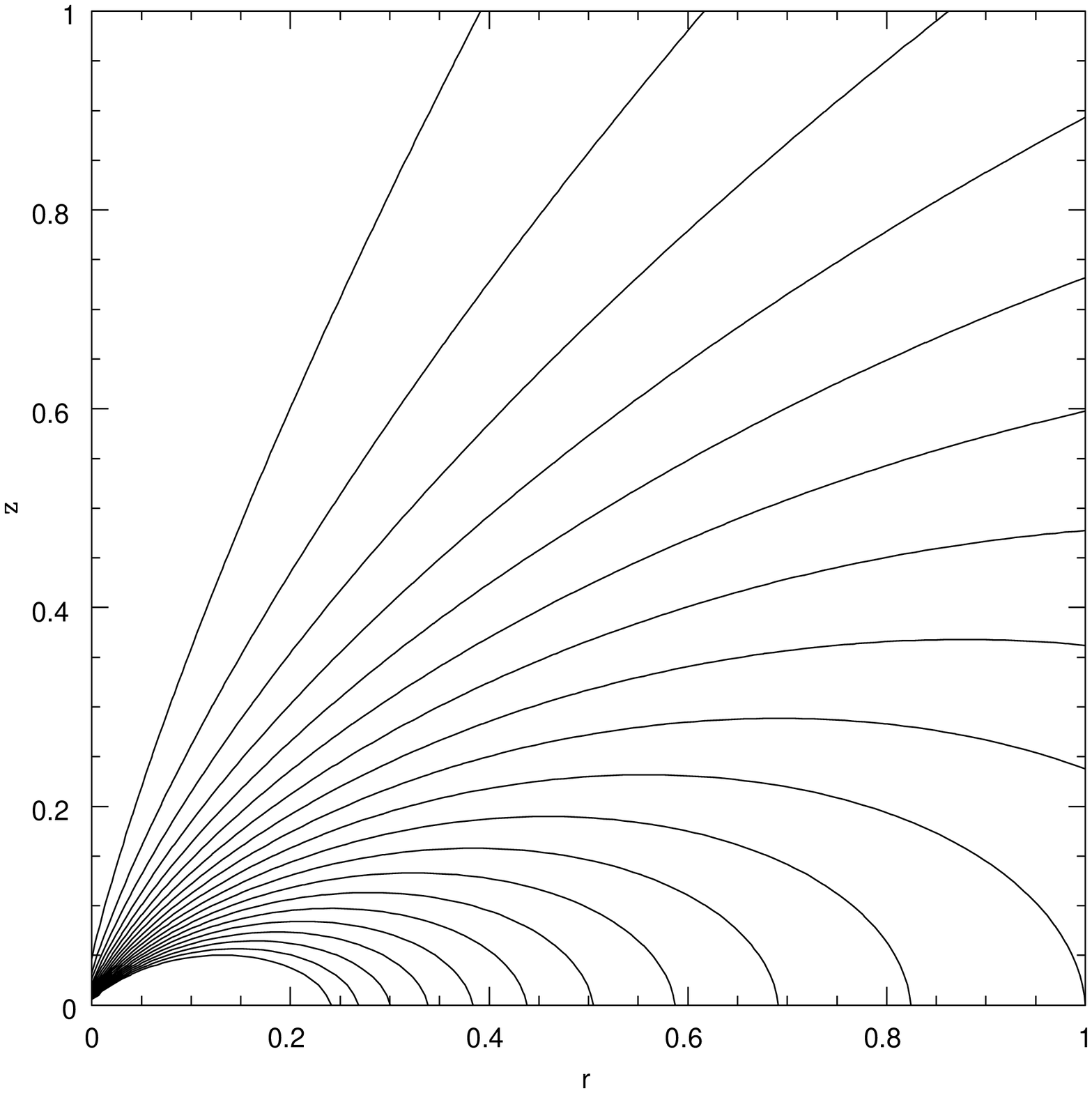}
    \includegraphics[angle=0, width=.4\textwidth]{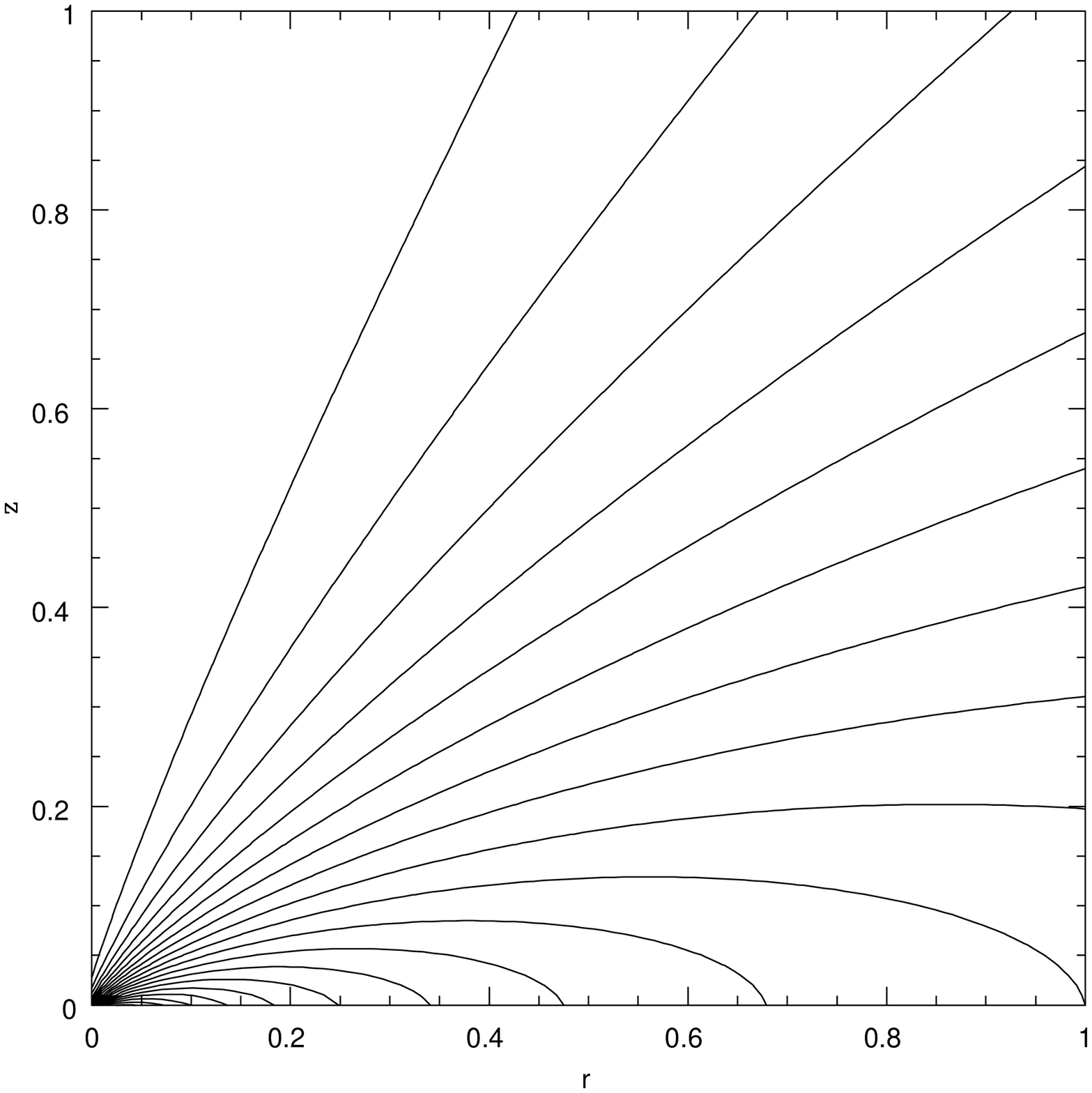}
    \caption{$\psi$-isolines of non-rotating star with poloidal current: (1) current $I\propto \psi ^3$, giving $\psi \propto R^{-1/2}$, maximal twist $118^\circ$ (2) current $I\propto \psi ^5$, giving $\psi \propto R^{-1/4}$, maximal twist $147^\circ$}
  \end{center}
\end{figure}

\end{document}